\documentclass[journal,10pt]{IEEEtran}
\usepackage{cite}
\usepackage{balance}
\usepackage{color}
\usepackage{epsfig}
\usepackage{epstopdf}
\usepackage{graphicx}
\usepackage{tabularx}
\usepackage{booktabs}

\usepackage{varwidth}

\usepackage{amssymb,amsmath,amsthm}

\usepackage{bm}
\usepackage{algorithm}%

\newfloat{routine}{htbp}{loa}
\floatname{routine}{Routine}
\makeatletter\newcommand\l@subroutine{\@dottedtocline{1}{1.5em}{2.3em}}\makeatother

\usepackage{algpseudocode}
\usepackage{pict2e}
\makeatletter
\def\BState{\State\hskip-\ALG@thistlm}

\makeatother

\newcommand{\m}[1]{\mathcal{#1}}

\usepackage{colortbl}

\newcounter{remarkCounter}

\newcounter{probCounter}

\begin{document}
\title{Cooperative Hierarchical Caching in 5G Cloud Radio Access Networks (C-RANs)}

\author{Tuyen~X.~Tran,~\IEEEmembership{Student Member,~IEEE,} Abolfazl Hajisami,~\IEEEmembership{Student Member,~IEEE,}  and
Dario~Pompili,~\IEEEmembership{Senior Member,~IEEE}
}

\markboth{IEEE Network, Feature Topic on Ultra-dense Heterogeneous Small Cell Deployment in 5G and Beyond, July~2017}
{Shell \MakeLowercase{\textit{et al.}}: Bare Demo of IEEEtran.cls for Computer Society Journals}

\maketitle

\thispagestyle{empty}\pagenumbering{arabic}
\IEEEdisplaynotcompsoctitleabstractindextext
\IEEEpeerreviewmaketitle

\begin{abstract}
Over the last few years, Cloud Radio Access Network~(C-RAN) is proposed as a transformative architecture for 5G cellular networks that brings the flexibility and agility of cloud computing to wireless communications. At the same time, content caching in wireless networks has become an essential solution to lower the content-access latency and backhaul traffic loading, leading into user Quality of Experience (QoE) improvement and network cost reduction. In this article, a novel Cooperative Hierarchical Caching~(CHC) framework in C-RAN is introduced where contents are jointly cached at the BaseBand Unit~(BBU) and at the Radio Remote Heads~(RRHs). Unlike in traditional approaches, the cache at the BBU, \emph{cloud cache}, presents a new layer in the cache hierarchy, bridging the latency/capacity gap between the traditional edge-based and core-based caching schemes. Trace-driven simulations reveal that CHC yields up to $51\%$ improvement in cache hit ratio, $11\%$ decrease in average content-access latency, and $18\%$ reduction in backhaul traffic load compared to the edge-only caching scheme with the same total cache capacity. Before closing the article, we discuss the key challenges and promising opportunities for deploying content caching in C-RAN in order to make it an enabler technology in 5G ultra-dense systems.
\end{abstract}

\section{Introduction}
Over the last few years, the proliferation of personal mobile devices such as smartphones and tablets, along with the plethora of Over-The-Top~(OTT) multimedia content providers (e.g., YouTube, Netflix, and Amazon) has resulted in an exponential growth in capacity demand in mobile wireless systems~\cite{cisco2012cisco}. Moreover, the future video encoding and playback advances (e.g., 4K resolution, very high quality encoding, and multi-angle) will further increase the capacity requirements. The demand in wireless networks has shifted from traditional connection-centric communications, such as phone calls and text messages, to multimedia content-centric communications such as video streaming and content sharing. While several solutions have been proposed to improve network capacity such as the deployment of ultra-dense small cells and massive MIMO, these approaches are fundamentally constrained by the limited spectrum resources and control signaling overhead. Therefore, in order to support the foreseen massive traffic in 5G networks in an affordable way, improving network capacity alone is not sufficient and has to be accompanied with innovations at higher layers (e.g., network architecture, backhaul transportation, and applications).

\begin{figure}
 \centering
 \includegraphics[scale = .5]{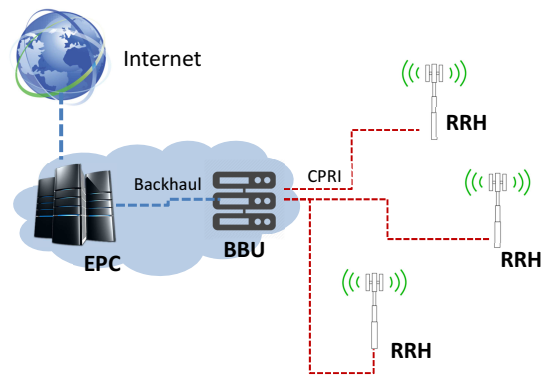}
\caption{C-RAN architecture with distributed RRHs connected to a common BBU via CPRI links. }\label{fig:cran}
\end{figure}

Recently, Cloud Radio Access Network~(C-RAN) has been introduced as a clean-slate redesign of cellular network architecture that addresses the capacity and coverage issues, while reducing operational costs and improving network flexibility~\cite{pompili2016elastic}. Using virtualization technologies such as Software-Defined Networking~(SDN) and Network Function Virtualization~(NFV), C-RAN shows also great potential in supporting autonomic network management (self-optimization, self-adaptation, and self configuration). In C-RAN, the computational functionalities are decoupled from the distributed Base Stations~(BSs) and consolidated in a centralized processing center. A typical C-RAN, as shown in Fig.~\ref{fig:cran}, is constituted of: (i) light-weight distributed Radio Remote Heads~(RRHs) deployed at the cell sites, (ii) a central BaseBand Unit~(BBU) pool hosted in a cloud datacenter, and (iii) high-bandwidth, low-latency fronthaul links connecting the RRHs to the BBU pool. The centralized nature of C-RAN along with virtualization technology enables dynamic resource allocation, mobility management and cooperative communications~\cite{tran2017twireless}. The cloud infrastructure at the BBU pool with strong computing resources and storage capacity now provides a central port for traffic offloading and content management to handle the growing Internet traffic from mobile users. This directly translates into Capital Expenditure~(CAPEX) and Operational Expenditure~(OPEX) reduction as well as user Quality of Experience~(QoE) improvement.

At the same time, in today's mobile networks, caching and computing resources are already ubiquitous, both at the BSs and on the user devices. Now, the fundamental question is how to effectively utilize the existing resources in order to address the need for massive content distribution. One promising approach is to enable content caching within the wireless operators' networks, where popular contents are cached at the BSs at the edge of the RAN. Upon receiving a content request from a user, the BS can provide the content that has been cached rather than downloading it from the original server. The benefits of content caching in cellular network have been explored recently in~\cite{bastug2014living, ahlehagh2014video, golrezaei2012femtocaching}. In particular, these works propose to alleviate backhaul usage via proactive caching at the small cell BSs, whereby files are proactively cached during off-peak hours based on file popularity and correlations among user and file patterns. In~\cite{Gharaibeh2015online}, the author also consider collaborative caching and introduce an online algorithm, that does not require prior knowledge about the content popularities, to minimize the total cost paid by content providers. More recently, the work in~\cite{li2016cooperative} proposes a cooperative RAN caching framework for coordinated multi-point joint transmission and single cell transmission based on a local altruistic game model that is aimed at minimizing content transmission time for mobile users. Huang et al.~\cite{huang2016content} study the content caching and user scheduling scheme for the heterogeneous networks and propose an algorithm to maximize the number of successfully scheduled users with limited radio resources. In the context of C-RAN, the works in~\cite{tao2016content, Mosleh2016Globecom} investigate the content caching and beamforming design to minimize the transmission power and backhaul cost. However, they only consider caching at the BS level and there is no collaboration among these caches.

Implementing caching at the edge of the network offers a significant backhaul traffic reduction. However, the aforementioned \emph{edge-only} caching schemes face two important drawbacks: (i) \emph{high cache miss ratio due to limited cache size at the BSs (compared to the very large number of content files)} and (ii) \emph{lack of consideration of user mobility from one cell to the other}. To compensate for the relative small cache size at the BSs, the authors in~\cite{wang2014cache} consider caches both in the RAN edge and in the Evolved Packet Core~(EPC). While it is possible to implement relatively large cache size at the EPC to improve the cache hit ratio, fetching content from EPC to the BSs still undertakes considerable delay due to the involvement of multiple intermediate network components.

In this article, we leverage the C-RAN architecture and propose a Cooperative Hierarchical Caching~(CHC) scheme. In particular, this scheme allocates relatively larger caches at the BBU pool (\emph{cloud cache}) and supplementary smaller caches at the distributed RRHs (\emph{edge cache}), while keeping the total cache size in the network fixed. The deployment of cloud cache and edge caches are complementary and are managed centrally by a controller in the BBU pool. While the hierarchical caching paradigm has been studied in the context of Content Delivery Networks~(CDNs), their settings and constraints are much different from the caching system in cellular wireless networks. Firstly, the limited spectrum resources coupled with the dynamic wireless channels and user mobility will highly affect the strategy of cache placement and content delivery in wireless networks. Secondly, the storage capacity of RAN caches will be much smaller than that of the caches in the CDN, making the cache placement critically important. In the simulation results, we have shown that our proposed CHC caching algorithm significantly outperforms the counterpart algorithm proposed for CDN~\cite{borst2010distributed}.

The remainder of this article is organized as follows. In Sect.~\ref{sec:dist_core}, we envision the new C-RAN architecture to support 5G ultra-dense networks. In Sect.~\ref{sec:chc}, we introduce our proposed novel caching strategy in C-RAN and present illustrative results. In Sect.~\ref{sec:open_research}, we discuss the key challenges and open-research directions that call for further investigation. Finally, we draw our conclusions in Sect.~\ref{sec:conclusion}.

\section{New C-RAN Architecture to Support 5G Ultra-dense Networks} \label{sec:dist_core}
We envision here the essential evolutions in C-RAN to support the key features of 5G ultra-dense cellular networks (i.g., ultra-high radio speed, ultra-low latency, massive connectivity, and content distribution).

{\bf New Fronthaul Interface:}
In C-RAN, all the RRHs are connected to the BBU pool via the fronthaul links using Common Public Radio Interface~(CPRI), which is more widely adopted in the industry than Open Base Station Architecture Initiative~(OBSAI). To realize high-speed CPRI links, the common solution is to use direct fiber connections, which are very limited and expensive. For example, in a TD-LTE system with $20~\rm{MHz}$ bandwidth and $2$ antennas on each RRH ($2~\rm{Tx}$, $2~\rm{Rx}$), the CPRI data rate between the BBU pool and the RRH for each TD-LTE carrier transmission is as high as $2.45~\rm{Gbps}$\footnote{Refer to: Common Public Radio Interface~(CPRI) Specification V6.0, Ericsson, 2013}. When considering a network with two bands and three RRHs (3-sector) on each cell site, the required fronthaul capacity is $14.7~\rm{Gbps}$. Suppose that each RRH uses one fiber link, then each cell site would require six fiber links. This number could be even higher in the more widely used FDD-LTE system where both downlink and uplink use $20~\rm{MHz}$ bandwidth. The high requirements for fronthaul fiber links makes it very costly and difficult to achieve for most operators due to the limited fiber resources. In addition, it usually takes long time to install fiber and in some locations it is not possible to install fiber at all. Once massive MIMO is deployed, fronthaul capacity between BBU and each RRH needs to be dramatically increased. Moreover, as new Radio Access Technologies~(RATs) are introduced, the bandwidth will extend from $20$ to beyond $100$ and $400~\rm{MHz}$, and so on, eventually requiring tens or hundreds of $\rm{Gbps}$ of CPRI capacity per RRH. The current C-RAN fronthaul architecture  with the maximum transmission capacity per fronthaul link of $10~\rm{Gbps}$ will no longer be able to handle such humongous capacity requirement of the 5G network. To overcome this issue, various solutions have been proposed, including new compression techniques, new transport modes for fronthaul transmission such as Wavelength-Division Multiplexing~(WDM) and microwave transmission.

\begin{figure}
 \centering
 \includegraphics[scale = .55]{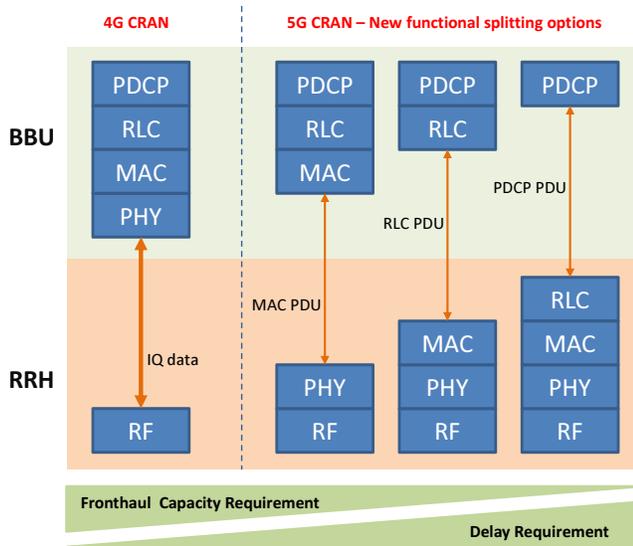}
\caption{Functional split between BBU and RRH in 4G and 5G C-RANs. Shifting more functionalities to the RRH decreases capacity requirement and increases delay requirement on the fronthaul links.
}\label{fig:split}
\end{figure}

{\bf Flexible Functional Splitting:}
While the alternative fronthaul solutions help reducing fiber consumption, a multi-fold reduction requires an architectural solution. Such transformative solution should redefine the functionalities of BBU pool and RRH differently from the current definition (where both the PHY and MAC are implemented in the BBU pool) and change the interface between BBU and RRH from Circuit fronthaul~(CPRI) to Packet fronthaul~(Ethernet). To make these changes, many functional split options have been proposed, each offering different trade-offs such as reduced fronthaul capacity and delay requirement. Figure~\ref{fig:split} illustrates the functional split between BBU pool and RRH and compares the 4G C-RAN with the new options in 5G C-RAN.

{\bf Distributed Core:}
In the 5G era, radio IP capacity will become as large as $20~\rm{Gbps}$ per sector and ultra-large content traffic--e.g., UHD video streaming, Augmented Reality~(AR), Virtual Reality~(VR)--will travel across the faster radio network. All mobile communication traffic has to travel via packet core network, i.e., Packet Gateway~(PGW). In current architecture, most countries have only a few sites with PGWs across their nations. If this current architecture is kept, massive backhaul between the BBU pools located across the country and packet core in a few centralized sites becomes inevitable and substantial backhaul investment has to be made as well. For instance, assuming there are 10,000 5G cell sites nationwide, each with 3 sectors and only half of the cell sites are at peak-traffic at the same time, the backhaul capacity required will be at least $300~\rm{Tbps}$ ($20~\rm{Gbps/sector}$ $\times$ 3-sector $\times$ 5,000 sites). Obviously, 5G core in centralized sites should have ultra high processing capacity as well. Because of the foregoing circumstances, it is highly desirable to have 5G solutions that involve distribution of 5G core nodes close to cell sites. In this case, content servers (or caching servers) can be placed on the rack right next to the distributed 5G core. This significantly helps reducing backhaul traffic by having mobile devices download content directly from the in-network content servers without having to pass the backhaul to each 5G core. In this scenario, 5G Core (Data Plane), BBU, and applications will run on virtualized servers at the local C-RAN sites.

\section{Cooperative Hierarchical Caching in C-RAN} \label{sec:chc}
In this section, we investigate the problem of cache placement in C-RAN aiming at minimizing the average content access delay of all users, and at the same time reducing backhaul traffic load. We consider a C-RAN that consists of $R$ RRHs distributed in $R$ corresponding cells, and a set of $U$ active users. The collection of files available for downloads is $\m{F} = \{f_1,f_2,...,f_F\}$. It is assumed that the popularities ranking of the files, $\m{P} = \{p_1, p_2,...,p_F\}$, are known \emph{a priori}. While predicting content popularity is a challenging task in terms of accuracy and scalability, the recent advances in machine learning and data mining techniques have made significant progress on achieving this goal. Such techniques could involve analyzing data from popular web sites, newspapers, and social networks in order to determine---around a specific RRH---what kind of content people like and search for, and what is the consumer's profile of these people. We envision that each RRH integrates not only the front Radio Frequency~(RF) but also certain capabilities to enable caching such as content storage and look up. This is similar to the idea of the evolved RRH proposed for Fog-based RAN~\cite{peng2016fog}. We consider that each RRH is equipped with a cache, $\m{C}_r$, with capacity $M_r~\rm[TB]$, while we assume that the BBU cloud has a larger cache with capacity $M_c~\rm[TB]$. Given that the storage capacity in each cache is limited, it is imperative to design an effective caching strategy that decides when, where, and what to store in each cache, so as to optimize the QoE of all users.

In current 4G cellular network, the eNodeBs are inter-connected via the X2 interface, which is designed for exchanging control information or user's data buffer during handover. While this X2 interface is available for limited data transfer, it cannot be exploited for inter-cache data transfer and hence the eNodeBs cannot share their cache contents directly. In contrast, the RRHs in C-RAN are all connected to the common BBU via high-bandwidth, low-latency CPRI links for user data transportation. This allows each RRH to retrieve cache contents from the neighboring RRHs via a ``U-turn'' (RRH-BBU-RRH) using fronthaul links. Note that retrieving cache data from neighboring RRHs is more latency- and cost-effective than fetching content from the original remote server via the backhaul network~\cite{wang2014cache, Gharaibeh2015online}. In this article, our proposed CHC strategy will fully exploit the extra degrees of cooperation brought by C-RAN to pool the resources and increase cache hit ratio, reducing outbound requests to the higher-level network elements.

\begin{figure}
 \centering
 \includegraphics[scale = .55]{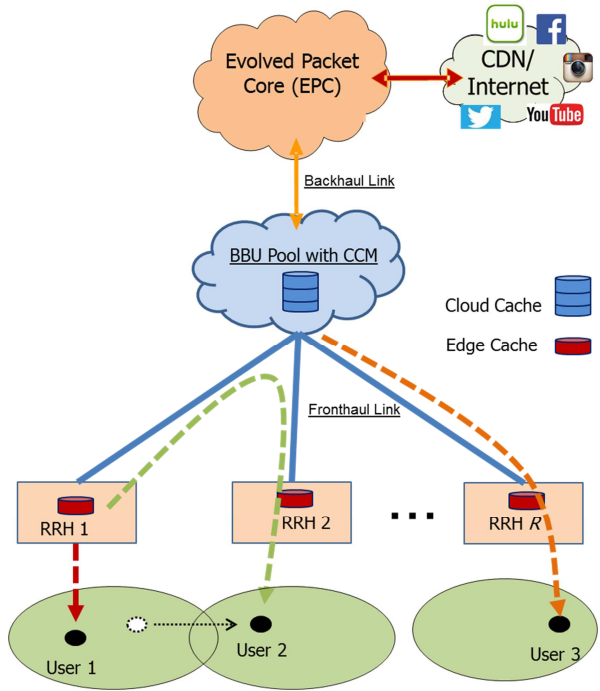}
\caption{Abstraction of cooperative hierarchical caching system in C-RAN.
}\label{fig:logical}
\end{figure}

{\bf Proposed Caching Strategy:}
In the proposed system, we consider that there is a Central Cache Manager~(CCM) implemented at the BBU cloud to monitor all the requests generated from users within the local C-RAN, which is responsible to make cache placement decision. In addition, leveraging the powerful processing capability at the BBU cloud, we can implement sophisticated learning and prediction algorithms to estimate the content popularity information $\m{P}$. While the actual content files are physically stored in the separated caches, a global indexing table can be maintained by the CCM to facilitate content lookup and cache management. A request for file $f$ from a user in cell $k$ will firstly go to the CCM. If the CCM determines that file $f$ is already cached at RRH $k$, it will inform RRH $k$ to send file $f$ directly to the user without incurring fronthaul traffic overhead. Otherwise, the CCM will search for $f$ in the cloud cache and in all the neighboring edge caches. If CCM cannot locate the requested file in any cache, it will direct the request to the original content server in the remote CDN, incurring traffic in the backhaul links. In Fig.~\ref{fig:logical}, we illustrate the abstraction of CHC system in C-RAN with an example where: a request from user~1 (in cell~1) and from user~2 (moved from cell~1 to cell~2) are retrieved from RRH~1's cache; and a request from user~3 in cell~R is retrieved from the cloud cache.

In the following, we introduce the cost model to characterize the cache management strategy. While the costs associated with access delay and bandwidth consumption are often proportional and interchangeable, we will focus on the cost model for content access delay (as considered in~\cite{golrezaei2012femtocaching} and \cite{wang2014cache}), which can be directly translated into user's QoE. Let $d_r$ denote the delay cost of transferring a file from BBU pool to RRH $r$, which we assume to be the same as the cost of retrieving that file from RRH $r$ in the uplink to the BBU. Let $d_0$ denote the delay cost incurred when a user in cell $r$ downloads a file from the original server in the CDN. Furthermore, we assume the cost of transferring a file from cache of RRH $k$ to RRH $r$ is $d_{rk} = d_r + d_k$. In practice, $d_0$ is usually many-fold higher than $d_r$ and $d_{rk}$~\cite{wang2014cache}. It should be noted that the cost of transferring files from the RRHs to the users always incur no matter whether caching is used or not, and it does not depend on the cache placement. Therefore, without loss of generality, we consider that the associated cost of a user downloading a file directly from the local cache of its serving RRH to be zero when considering cache placement design.
Hence, the average delay cost of user $u$ in cell $r$ can be expressed as,
\begin{equation} \label{eq:delay_u}
{\bar{D}_u} = \sum\limits_{i = 1}^F {{p_i}\left( {x_{ir}^0{d_r} + \sum\limits_{k = 1,k \ne r}^R {x_{ir}^k{d_{rk}}}  + x_{ir}^{R + 1}{d_0}} \right)},
\end{equation}where $x_{ir}^k = 1$ if request for file $f_i$ from user $u$ is retrieved from cache $\m{C}_r$ and  $x_{ir}^k = 0$ otherwise ($k=0,1,...,R$), and $x_{ir}^{R+1}$ indicates whether such request is retrieved from the CDN. The optimal cache placement algorithm can be obtained by solving the optimization problem that minimizes the total average delay of all users in the network, $\sum\limits_{u = 1}^U {{{\bar D}_u}}$, subject to the set of cache capacity constraints. In general, the cache placement problem is NP-complete, which is impractical to implement due to the high-complexity solution. Equivalently, we can recast the cache-placement problem as the problem of maximizing a system utility function---the expected delay cost saving---and show that this problem belongs to the class of maximizing a monotone submodular set function over a matroid constraint. Our proposed greedy CHC strategy comprises of two phases, namely the \emph{Proactive Cache Distribution}~(PCD) phase and the \emph{Reactive Cache Replacement}~(RCR) phase. \emph{Firstly}, the PCD phase involves building a cache placement solution starting with empty caches and incrementally adding files to the caches. In each iteration, it adds a new file with the highest marginal value to the cache placement set, until all the caches are full. Since the objective function is submodular, the marginal value of a new file decreases as the cache placement set grows bigger. Furthermore, due to the monotonicity of the objective function, the PCD algorithm is guaranteed to achieve an objective function of at least $\frac{1}{2}$ the optimal value~\cite{nemhauser1978analysis}. This phase can be done during off-peak traffic hours (e.g., night time) to utilize the unused backhaul bandwidth. \emph{Secondly}, the RCR phase occurs over the course of the day to make cache-replacement decision. In particular, following each cache miss, a new file will be downloaded from the remote content server to the local RAN and delivered to the requesting user. The RCR algorithm will decide to replace this new file with existing files in the cache only if such replacement could improve the value of the objective function. 

{\bf Performance Evaluation:}
We carry out trace-driven simulations to evaluate the performance of the aforementioned CHC strategy in terms of cache hit ratio, average latency, and backhaul traffic load. We use the YouTube request trace data collected on the campus of University of Massachusetts Amherst  during the day $03/12/2008$\footnote{Refer to: http://traces.cs.umass.edu/index.php/Network/Network}. The video popularities are extracted from the trace and are used as the input for the greedy CHC algorithm.
Based on the video request data, we simulated a C-RAN system with each cell having one RRH and mobile users are uniformly distributed in the cells. The e2e latency of video delivery from the CDN to RRH and from BBU cloud to RRH are assigned to be $d_0 = 100~\mathrm{ms}$ and $d_{r} = 20~\mathrm{ms}$, respectively. We assume that the backhaul and fronthaul link capacity as well as radio resources in the access network are sufficiently provisioned to handle all the generated traffic requests. We allocate the cache capacity in the BBU cloud to be four times larger than cache capacity in each RRH, i.e., ${M_0} = 4{M_r}$, while keeping the total cache capacity in the network the same for all considered scenarios. The backhaul traffic volume is calculated as the amount of traffic downloading from the CDN due to cache misses. The trace-driven simulation parameters are summarized in Table~\ref{tab:para}.

\begin{table}[ht]
\caption{Trace-Driven Simulation Parameters \label{tab:para}}
\begin{center}
\begin{tabular}{lc}
\hline \bottomrule \hline \\ [-1.5ex]
Description & Value \\ [.5ex]
\hline \\[-1.5ex]
Number of Cells & $4$ \\ [.5ex]
Number of Users & $19,777$ \\ [.5ex]
User Distribution & $\mathrm{Uniform}$ \\ [.5ex]
Number of Videos & $77,414$ \\ [.5ex]
Video Size & $20~\mathrm{MB}$\\ [.5ex]
Number of Request & $122,280$	\\ [.5ex]
BBU-RRH Latency & $20~\mathrm{ms}$ \\ [.5ex]
CDN-RRH Latency & $100~\mathrm{ms}$ \\  [1ex]
 \bottomrule
\end{tabular}
\end{center}
\end{table}

\begin{figure*}[t]
 \centering
 \begin{tabular}{ccc}
\hspace*{-.3cm}\includegraphics[scale = .6]{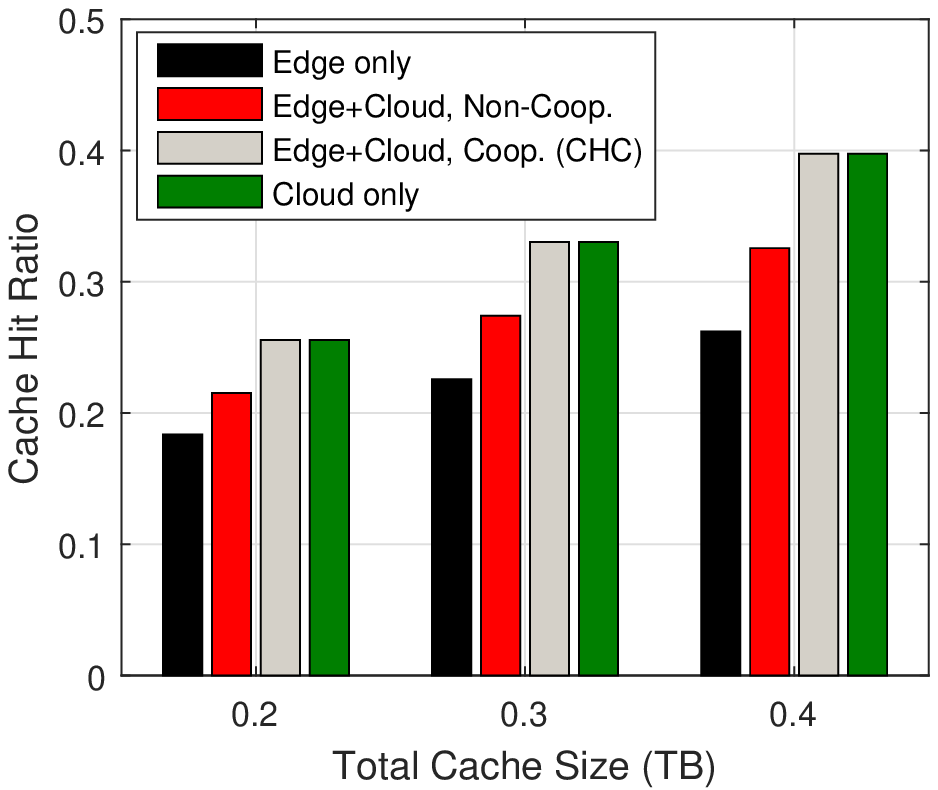} &
\hspace*{-.6cm}\includegraphics[scale = .6]{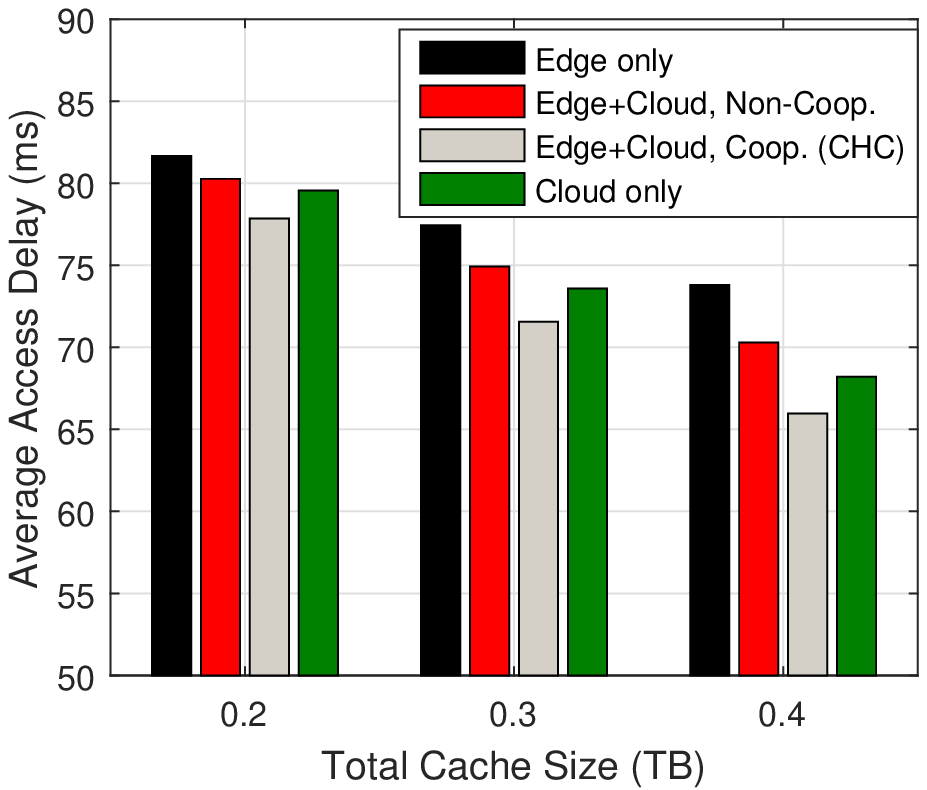} &
\hspace*{-.6cm}\includegraphics[scale = .6]{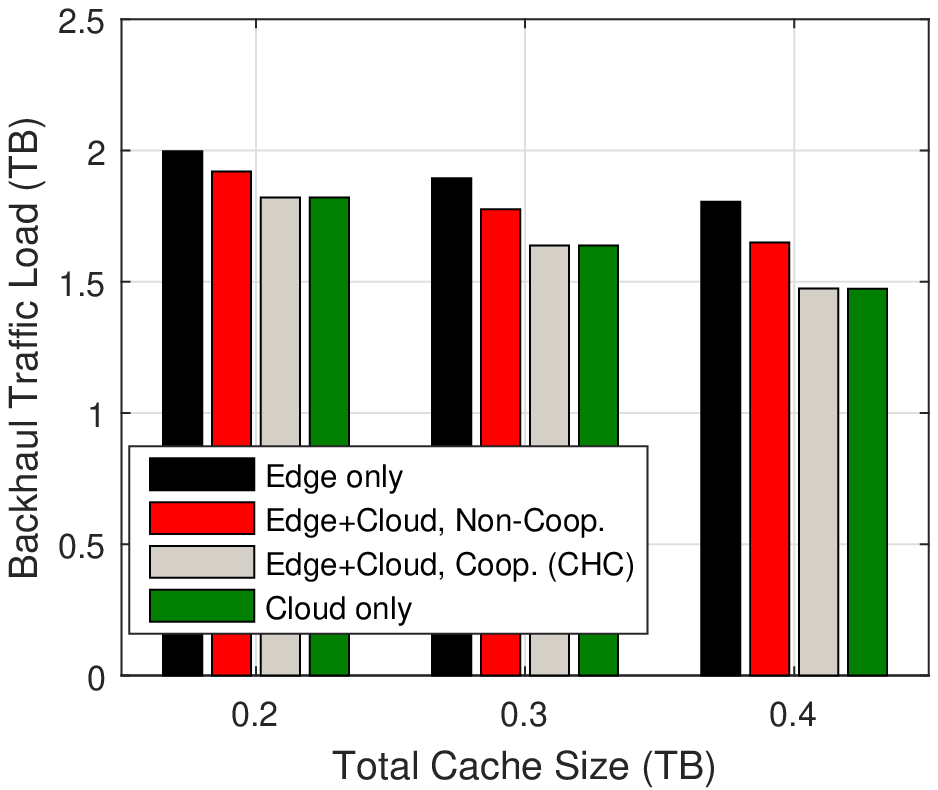} \\
 \small(a) & \small(b) & \small(c)
\end{tabular}
\caption{Performance of different caching strategies in a C-RAN are evaluated over different metrics in terms of different total cache sizes: (a) Cache Hit Ratio; (b) Average Latency; (c) Backhaul Traffic Load.}
\label{fig:cache_architecture}
\end{figure*}

{\emph{\underline{Impact of caching architecture:}}}
We compare the performance of four caching architectures.
\begin{itemize}
\item \emph{Edge only}: popular files are cached at the RRHs only. If the requested file from a mobile user is found in its home RRH's cache, the file will be download immediately from the cache; otherwise, it will be fetched from the CDN.

\item \emph{Edge+Cloud, Non-Coop.}: a hierarchical caching strategy where contents are cached at both the BBU and the RRHs, however there is no cooperation among the RRHs' caches. A file request resulting in a cache miss in the RRH will be searched in the BBU cloud's cache and finally goes out to the CDN.

\item \emph{Edge+Cloud, Coop.~(CHC)}: our proposed cooperative hierarchical caching strategy.

\item \emph{Cloud only}: contents are cached at the BBU cloud only. 

\end{itemize}

From Figs.~\ref{fig:cache_architecture}(a-c), we see that \emph{for the same total cache capacity}, allocating caches at both the BBU cloud and RRHs provides significant performance improvement compared to allocating caches at the RRH only. For example, the performance gains of CHC scheme over the \emph{Edge only} scheme when the total cache size is $0.4~\rm{TB}$ are approximately $51\%$ improvement in cache hit ratio, $11\%$ decrease in average e2e latency, and $18\%$ reduction in backhaul traffic load. In addition, the performance is further improved by cooperating the RRH's caches, which are characterized by the gains of CHC scheme over the \emph{Non-Coop.} scheme. The cache hit performance and backhaul usage of the CHC scheme and the \emph{Cloud only} scheme are almost the same; however, CHC achieves significantly lower average access delay.

\begin{figure*}[t]
 \centering
 \begin{tabular}{ccc}
\hspace*{-.3cm}\includegraphics[scale = .6]{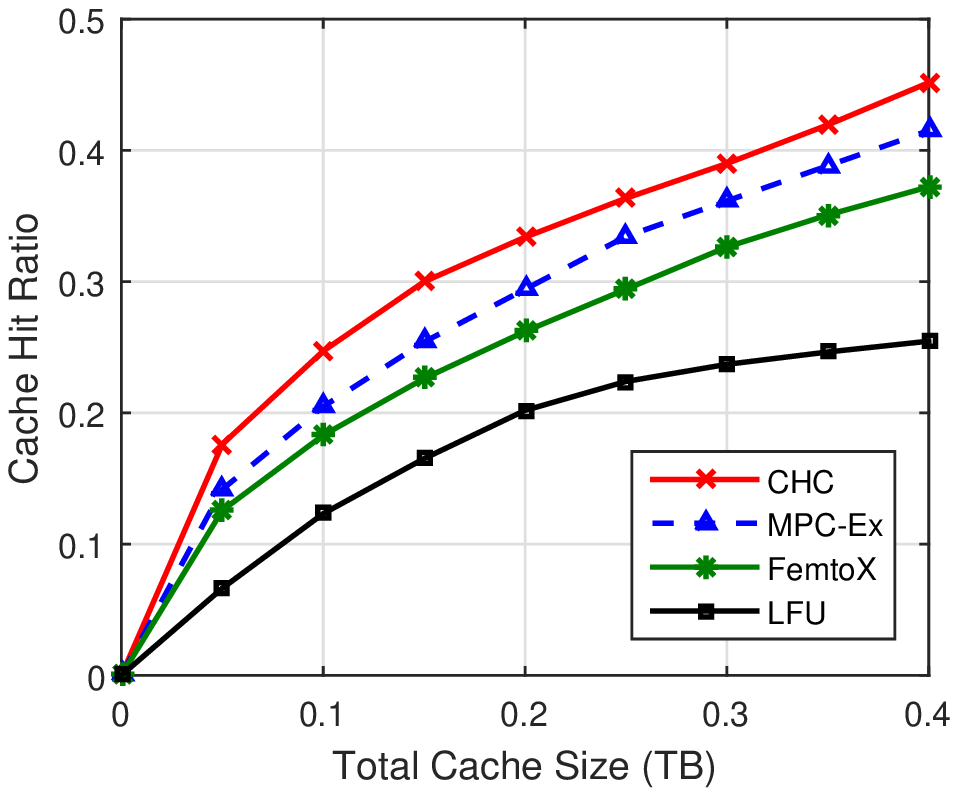} &
\hspace*{-.6cm}\includegraphics[scale = .6]{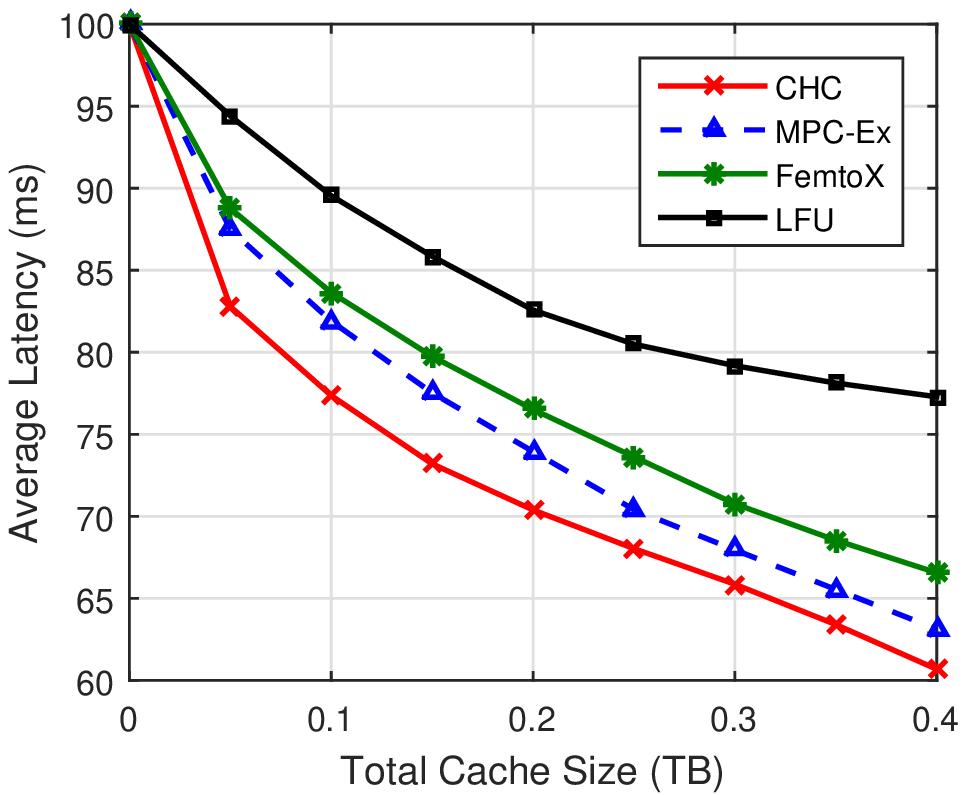} &
\hspace*{-.6cm}\includegraphics[scale = .6]{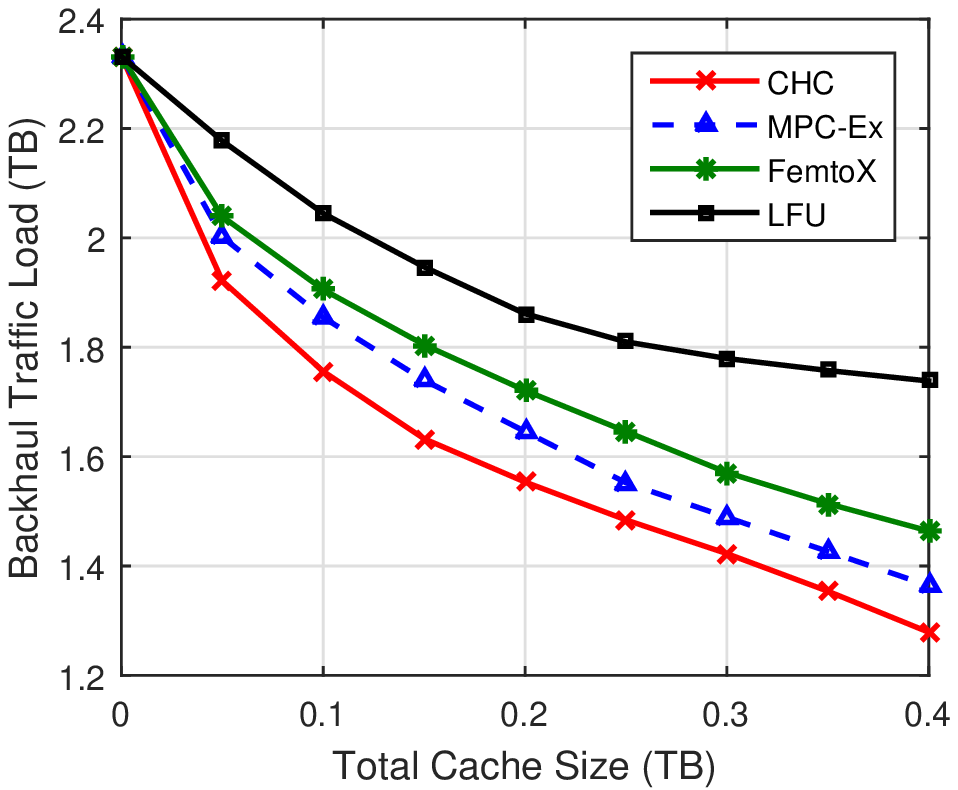} \\
 \small(a) & \small(b) & \small(c)
\end{tabular}
\caption{Performance of our Cooperative Hierarchical Caching (CHC) with different cache replacement algorithms: (a) Cache Hit Ratio; (b) Average Latency; (c) Backhaul Traffic Load. 
}\label{fig:cache_replacement}
\end{figure*}

{\emph{\underline{Impact of cache replacement policy:}}}
To evaluate the impact of different caching policies in C-RAN, we compare the proposed CHC strategy with three baselines: (i) MPC-Ex: the Most Popular Caching scheme where each cache independently stores the most popular files and the files that are stored in the edge caches are excluded from the cloud cache~\cite{borst2010distributed}; (ii) FemtoX: an extension of the FemtoCaching scheme~\cite{golrezaei2012femtocaching} to C-RAN; and (iii) LRU: the Least Recently Used cache replacement algorithm~\cite{jelenkovic2004least}. From Figs.~\ref{fig:cache_replacement}(a-c), we can see that the CHC scheme significantly outperforms the baselines in terms of cache hit ratio improvement as well as average latency and backhaul traffic loading reduction.

\section{Open Research Directions} \label{sec:open_research}
In this section, we discuss the key challenges and open-research issues that need to be addressed in order to fully exploit the benefits of content caching in 5G C-RAN.

{\bf On-device caching:}
With mobile devices now hosting multi gigabytes of main storage, the idea of buffering multiple videos on these devices is becoming more relevant. The first interesting problem will be the predictive part, i.e., how does that work? how does the system know what a user might want to watch? The prediction involves interrogating user preference and content-recommendation engines so that the personalized video content can be \emph{pre-position} to the mobile devices. This mechanism may work as a personalized catch-up TV service, falling between ``video on demand'' and ``broadcast channel''. On the other hand, once on-device caches and the underlying device-to-device~(D2D) communications become popular, it is imperative to exploit the possible opportunistic coordination among these caches. In particular, the CCM in the BBU pool can explore the social relationships and ties among users to identify \emph{influential} users using the social graph. When a given user requests a file, the CCM determines whether one of the influential users has the requested file and, if so, direct the influential user to transfer the file to the requesting user via D2D. While this approach benefits the network and the requesting users, it is quite challenging, e.g., how to design an effective incentive model to attract users having caches to form the coalition is an open research problem.

{\bf Collaboration of OTT and network operator:}
To improve revenue per subscriber, OTT content providers want to expand their subscriber base and provide high-quality video service. This, in turn, increases the video traffic prompting the content provider to pay more for CDN services. By positioning the caching servers in C-RAN, network operators can deliver high-quality video contents to users  by utilizing idle bandwidth of the fronthaul links  without increasing load on the backbone network, securing  a new revenue model (caching fee) without additional  investment in their IP networks. At the same time, OTT providers can provide high quality service at a lower  cost than using the 3rd party CDN through the caching server in the operator networks. This helps expand their subscriber base and enhance customer QoE. A new business model will play a key role to achieve a win-win game between OTT providers and network operator on the share of caching resources.

{\bf Cross-layer design:}
To leverage the extra degrees of cooperation in C-RAN, many advanced cooperation techniques involving user scheduling, RRH clustering, and beamforming design have been proposed to optimize the spectral and/or energy efficiency. When cache storage are deployed on the RRHs, they will bring new optimization dimension to the existing approaches. For example, a user requesting a file might be scheduled to connect to the RRH having the requested file in the cache, instead of connecting to the closest RRH. It would be very interesting to design a holistic cooperation strategy in C-RAN taking into account both the  channel opportunities and the cache availability. 

\balance

\section{Conclusions} \label{sec:conclusion}
We introduced a novel Cooperative Hierarchical Caching~(CHC) framework in the context of Cloud Access Radio Networks~(C-RAN), where contents are jointly cached at the BaseBand Unit~(BBU) pool and at the Radio Remote Heads~(RRHs). The cloud cache in the BBU pool is envisioned as a new layer in the cache hierarchy that coordinates with the edge caches at the RRHs. Trace-driven simulations show that CHC significantly outperforms traditional edge-only caching scheme, rendering up to $51\%$ improvement in cache hit ratio, $11\%$ decrease in content-access latency, and $18\%$ reduction in backhaul traffic load. We also highlighted related challenges and opportunities of deploying content caching in C-RAN.
\\\textbf{Acknowledgment: }This work was supported by the National Science Foundation~(NSF) under Grant No.~CNS-1319945.

\bibliographystyle{ieeetr}\small

\begin{thebibliography}{10}
\providecommand{\url}[1]{#1}
\csname url@samestyle\endcsname
\providecommand{\newblock}{\relax}
\providecommand{\bibinfo}[2]{#2}
\providecommand{\BIBentrySTDinterwordspacing}{\spaceskip=0pt\relax}
\providecommand{\BIBentryALTinterwordstretchfactor}{4}
\providecommand{\BIBentryALTinterwordspacing}{\spaceskip=\fontdimen2\font plus
\BIBentryALTinterwordstretchfactor\fontdimen3\font minus
  \fontdimen4\font\relax}
\providecommand{\BIBforeignlanguage}[2]{{%
\expandafter\ifx\csname l@#1\endcsname\relax
\typeout{** WARNING: IEEEtran.bst: No hyphenation pattern has been}%
\typeout{** loaded for the language `#1'. Using the pattern for}%
\typeout{** the default language instead.}%
\else
\language=\csname l@#1\endcsname
\fi
#2}}
\providecommand{\BIBdecl}{\relax}
\BIBdecl
%
\bibitem{cisco2012cisco}
I.~Cisco, ``Cisco visual networking index: Forecast and methodology,
  2011--2016,'' {\em CISCO White paper}, 2012.

\bibitem{pompili2016elastic}
D.~Pompili, A.~Hajisami, and T.~X. Tran, ``{Elastic resource utilization
  framework for high capacity and energy efficiency in Cloud RAN},'' {\em IEEE
  Commun. Mag.}, vol.~54, no.~1, pp.~26--32, 2016.

\bibitem{tran2017twireless}
T.~X. Tran and D.~Pompili, ``{Dynamic Radio Cooperation for User-Centric
  Cloud-RAN with Computing Resource Sharing},'' {\em IEEE Trans. on Wireless
  Commun.}, 2017, to appear.

\bibitem{bastug2014living}
E.~Bastug, M.~Bennis, and M.~Debbah, ``{Living on the edge: The role of
  proactive caching in 5G wireless networks},'' {\em IEEE Commun. Mag.},
  vol.~52, no.~8, pp.~82--89, 2014.

\bibitem{ahlehagh2014video}
H.~Ahlehagh and S.~Dey, ``Video-aware scheduling and caching in the radio
  access network,'' {\em IEEE/ACM Trans. Netw.}, vol.~22, no.~5,
  pp.~1444--1462, 2014.

\bibitem{golrezaei2012femtocaching}
N.~Golrezaei, K.~Shanmugam, A.~G. Dimakis, A.~F. Molisch, and G.~Caire,
  ``Femtocaching: Wireless video content delivery through distributed caching
  helpers,'' in {\em Proc. IEEE INFOCOM}, pp.~1107--1115, 2012.

\bibitem{Gharaibeh2015online}
A.~Gharaibeh, A.~Khreishah, B.~Ji, and M.~Ayyash, ``A provably efficient online
  collaborative caching algorithm for multicell-coordinated systems,'' {\em
  IEEE Trans. Mobile Comput.}, vol.~PP, no.~99, pp.~1--14, 2015.

\bibitem{li2016cooperative}
H.~Li, C.~Yang, X.~Huang, N.~Ansari, and Z.~Wang, ``{Cooperative RAN Caching
  based on Local Altruistic Game for Single and Joint Transmissions},'' {\em
  IEEE Commun. Letters}, 2016.

\bibitem{huang2016content}
X.~Huang and N.~Ansari, ``Content caching and user scheduling in heterogeneous
  wireless networks,'' in {\em Proc. IEEE GLOBECOM}, pp.~1--6, 2016.

\bibitem{tao2016content}
M.~Tao, E.~Chen, H.~Zhou, and W.~Yu, ``Content-centric sparse multicast
  beamforming for cache-enabled cloud {RAN},'' {\em IEEE Trans. on Wireless
  Commun.}, vol.~15, no.~9, pp.~6118--6131, 2016.

\bibitem{Mosleh2016Globecom}
S.~Mosleh, L.~Liu, H.~Hou, and Y.~Yi, ``{Coordinated Data Assignment: A Novel
  Scheme for Big Data Over Cached Cloud-RAN},'' {\em Proc. IEEE GLOBECOM}, Dec.
  2016.

\bibitem{wang2014cache}
X.~Wang, M.~Chen, T.~Taleb, A.~Ksentini, and V.~Leung, ``{Cache in the air:
  exploiting content caching and delivery techniques for 5G systems},'' {\em
  IEEE Commun. Mag.}, vol.~52, no.~2, pp.~131--139, 2014.

\bibitem{borst2010distributed}
S.~Borst, V.~Gupt, and A.~Walid, ``Distributed caching algorithms for content
  distribution networks,'' in {\em Proc. IEEE INFOCOM}, pp.~1--9, 2010.

\bibitem{peng2016fog}
M.~Peng, S.~Yan, K.~Zhang, and C.~Wang, ``Fog-computing-based radio access
  networks: issues and challenges,'' {\em IEEE Network}, vol.~30, no.~4,
  pp.~46--53, 2016.

\bibitem{nemhauser1978analysis}
G.~L. Nemhauser, L.~A. Wolsey, and M.~L. Fisher, ``{An analysis of
  approximations for maximizing submodular set functions - I},'' {\em
  Mathematical Programming}, vol.~14, no.~1, pp.~265--294, 1978.

\bibitem{jelenkovic2004least}
P.~R. Jelenkovi{\'c} and A.~Radovanovi{\'c}, ``Least-recently-used caching with
  dependent requests,'' {\em Theoretical computer science}, vol.~326, no.~1,
  pp.~293--327, 2004.

\end{thebibliography}


\vspace{0.4cm}
\section*{Biography}\label{sec:bio}
\vspace{0.2cm}
\textbf{Tuyen X. Tran} is working towards his PhD degree in Electrical and Computer Engineering~(ECE) at Rutgers U. under the guidance of Dr.~Pompili. He received the MSc degree in ECE from the U. of Akron, USA, in 2013, and the BEng degree (Honors Program) in Electronics and Telecommunications from Hanoi U. of Technology, Vietnam, in 2011. His research interests are on the applications of optimization and machine learning to wireless networking systems. 

\vspace{0.2cm}
\textbf{Abolfazl Hajisami} started his PhD program in ECE at Rutgers U. in 2012. Currently, he is pursuing research in the fields of C-RAN, cellular networking, and mobility management under the guidance of Dr.~Pompili. Previously, he had received his MS and BS from Sharif U. of Technology and Shahid Beheshti U. (Tehran, Iran), in 2010 and 2008, respectively. His research interests are in wireless communications, C-RAN, statistical signal processing, and image processing.

\vspace{0.2cm}
\textbf{Dario Pompili} is an Assoc. Prof. with the Dept. of ECE  at Rutgers U., where he directs the Cyber-Physical Systems Laboratory~(CPS Lab). He received his PhD in ECE from the Georgia Institute of Technology in 2007. He had previously received his `Laurea' (combined BS and MS) and Doctorate degrees in Telecommunications and Systems Engineering from the U. of Rome ``La Sapienza,'' Italy, in 2001 and 2004, respectively. He is a recipient of the NSF CAREER'11, ONR Young Investigator Program'12, and DARPA Young Faculty'12 awards. He is a Senior Member of both the IEEE Communications Society and the ACM.

\end{document}